\documentclass[twocolumn,prb,amsfonts,amsmath,amssymb,floatfix,superscriptaddress]{revtex4} 
\usepackage{color}
\usepackage{hhline}
\usepackage{mathrsfs}
\usepackage{graphicx}
\usepackage{dcolumn}
\usepackage{bm}
\usepackage{multirow}
\usepackage{booktabs}
\usepackage{afterpage}
\usepackage{amsmath}
\usepackage[T1]{fontenc}

\begin{document}

\title{Theoretical analysis of the origin of the double-well band dispersion in the CuO double chains of Pr$_2$Ba$_4$Cu$_7$O$_{15-\delta}$ and its impact on superconductivity}

\author{Toshiki Yagi}
\affiliation{Department of Physics, Osaka University, Machikaneyama-cho, Toyonaka, Osaka 560-0043, Japan}

\author{Masayuki Ochi}
\affiliation{Department of Physics, Osaka University, Machikaneyama-cho, Toyonaka, Osaka 560-0043, Japan}
\affiliation{Forefront Research Center, Osaka University, Machikaneyama-cho, Toyonaka, Osaka 560-0043, Japan}

\author{Kazuhiko Kuroki}
\affiliation{Department of Physics, Osaka University, Machikaneyama-cho, Toyonaka, Osaka 560-0043, Japan}
\email{kuroki@presto.phys.sci.osaka-u.ac.jp}

\date{\today}
\begin{abstract}
Pr$_2$Ba$_4$Cu$_7$O$_{15-\delta}$ is a unique member of cuprate superconductors where many studies suggest that CuO double chains are responsible for superconductivity.
One characteristic and non-trivial feature of its electronic structure is a relatively large electron hopping $t$ between nearest neighbor Cu sites with a Cu-O-Cu angle of around 90 degrees.
In this study, we have theoretically pinned down the origin of a large $|t|$ in the double-chain structure of Pr$_2$Ba$_4$Cu$_7$O$_{15-\delta}$ using first-principles calculation and tight-binding-model analysis.
We have found that, in the nearest neighbor hopping $t$, $d$-$d$ and $d$-$p$-$p$-$d$ contributions roughly cancel each other out and the $d$-$p$-$d$ hopping path enhanced by the local distortion of the double chain is a key to get the large $|t|$. Double-well band dispersion arising from the relatively large $|t/t'|$ allows the enhancement of spin-fluctuation-mediated superconductivity by the incipient-band mechanism, where the one band bottom plays a role of the incipient valley. Our study provides the important knowledge to understand the unique superconductivity in Pr$_2$Ba$_4$Cu$_7$O$_{15-\delta}$.
\end{abstract}

\maketitle

\section{Introduction}

The well-known high $T_c$ family of the cuprate superconductors takes various crystal structures, but most of the family members possess CuO$_2$ planes playing the main role in the occurrence of superconductivity. Nonetheless, there are some exceptions: a well known example is (Sr,Ca)$_{14}$Cu$_{24}$O$_{41}$~\cite{Akimitsu}, in which superconductivity, under pressure, originates from the ladder-like structure within the compound. Another example is {\it Ae}$_2$CuO$_{3+\delta}$ ($Ae=$ Sr~\cite{Hiroi}, Ba~\cite{Uchida}), for which the crystal structure is yet to be determined completely. Yet another example is  Pr$_2$Ba$_4$Cu$_7$O$_{15-\delta}$ with $T_c$ ranging from 15 to 30 K~\cite{Matsukawa,Yamada,Honnami,Watanabe,Sasaki,Toshima}. This material consists of an alternating stack of PrBa$_2$Cu$_3$O$_{7-\delta}$, which consists of CuO$_2$ planes and CuO single chains with oxygen deficiencies, and PrBa$_2$Cu$_4$O$_8$ composed of CuO$_2$ planes and CuO double chains, where a number of studies suggest that the double chains are responsible for superconductivity. Experimentally, a Cu nuclear quadrupole resonance (NQR) study has shown that the reduction of $1/T_1T$ is seen below $T_c$ at the double-chain sites~\cite{Watanabe}, suggesting the opening of the superconducting gap at those sites. More recently, a Cu NQR study on a sample exhibiting 100\% superconducting volume fraction has shown that the CuO$_2$ planes are insulating and undergoes antiferromagnetic ordering at low temperatures~\cite{Sasaki}. 

\begin{figure}
\begin{center}
\includegraphics[width=8.5 cm]{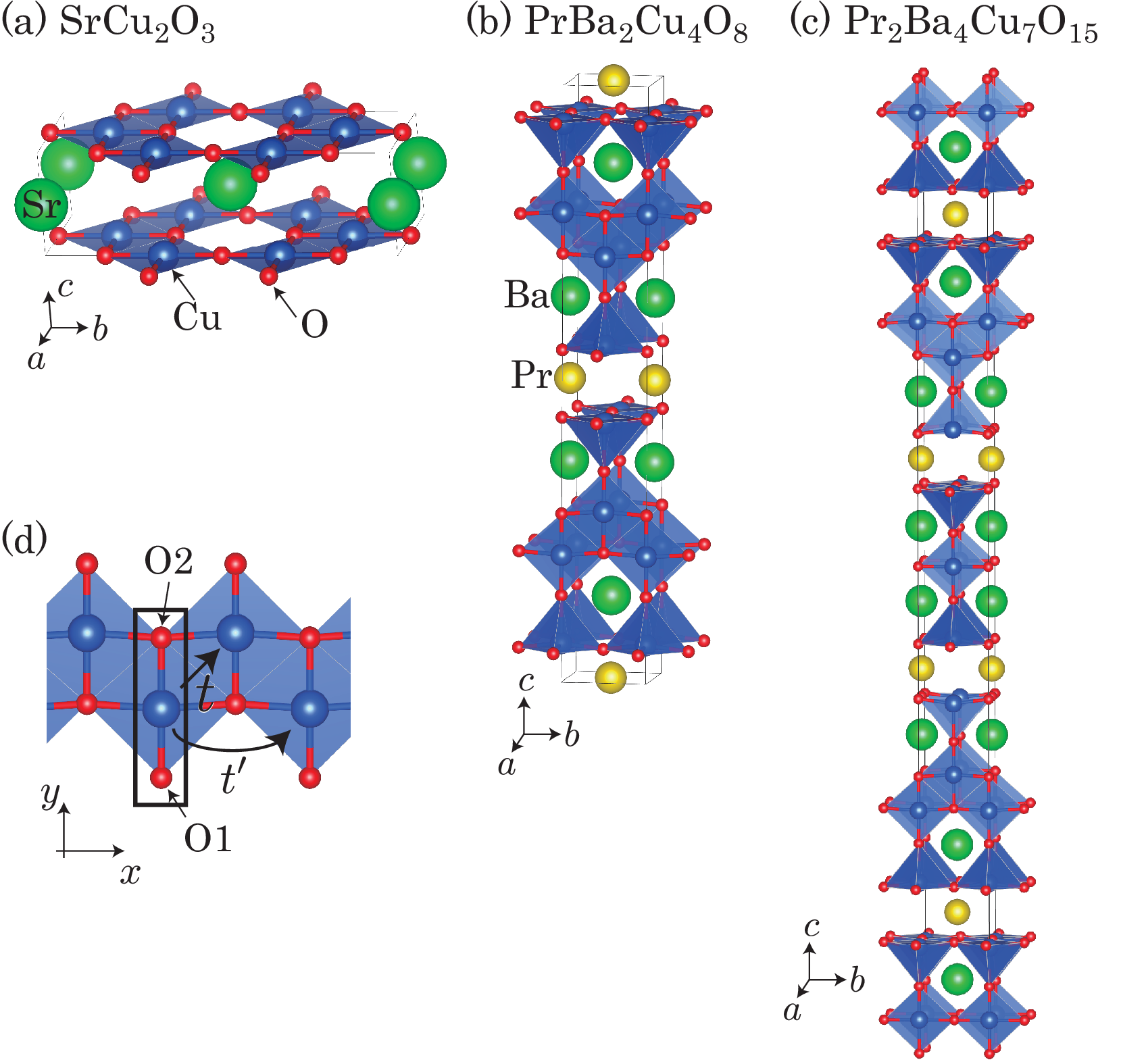}
\caption{Crystal structures of (a) SrCu$_2$O$_3$, (b) Pr$_2$Ba$_4$Cu$_7$O$_{15}$, and (c) PrBa$_2$Cu$_4$O$_8$. (d) CuO$_2$ double-chain structure commonly existing in these three compounds. The nearest neighbor Cu-Cu hopping $t$ and the next nearest neighbor Cu-Cu hopping $t'$ are shown in (d). A black solid square shows the unit cell used in model calculations, where a neighboring unit cell can be represented with a glide reflection. Depicted using the VESTA software~\cite{VESTA}.}
\label{fig:strct}
\end{center}
\end{figure}

Theoretically, various authors have studied the possibility of unconventional superconductivity taking place in the double chain structure~\cite{Sano,Nakano,Sano2,Okunishi,Berg,Nishimoto,Habaguchi,Kaneko}. From a tight-binding point of view, much focus has been paid on the electron hopping $t$ between nearest neighbor Cu sites, which reside on different chains among the double-chain structure [see Fig.~\ref{fig:strct}(d)]. This type of electron hopping is generally much smaller compared to the next nearest neighbor hopping $t'$ within the chains because the Cu-O-Cu path in the former forms an angle of $\sim$ 90 degrees, while the angle is $\sim$ 180 degrees in the latter. The effect of finite $|t/t'|$ on the band structure can be understood more easily by unfolding the Brillouin zone (i.e. by taking a reduced unit cell that contains one Cu site), which results in a single band with a double-well structure. Hence, for some range of the band filling, there exist four Fermi points. Considering the Hubbard model with such a band structure from a weak coupling viewpoint, in which only the band structure around the Fermi level matters, the system can be considered as equivalent to a two-leg Hubbard ladder system, which also exhibits four Fermi points~\cite{Fabrizio,KKArita}, and is known to exhibit superconductivity.  Such an effect is obviously absent for $t=0$, where the double-well structure of the band is lost (in the original double-chain structure, the interchain coupling is lost), so that the effect is naturally expected to be weak when $|t/t'|\ll 1$. Quite recently, this problem has been addressed from a strong coupling point of view, where a one-dimensional $t$-$J$ model with nearest neighbor $J$ and next nearest neighbor $J'$ were investigated using density matrix renormalization group technique~\cite{Kaneko}. There also, antiferromagnetic nearest neighbor $J$, which can be considered as originating from $t$ in the strong on-site $U$ limit, is required for the occurrence of superconductivity. In fact, our analysis in the present study shows that $|t/t'|$ is relatively large in Pr$_2$Ba$_4$Cu$_7$O$_{15}$, compared to, e.g., a well-known two-leg ladder cuprate SrCu$_2$O$_3$, where the two neighboring chains of adjacent ladders form a similar lattice structure as in the double chains of Pr$_2$Ba$_4$Cu$_7$O$_{15}$. The origin of the relatively large $|t/t'|$ in Pr$_2$Ba$_4$Cu$_7$O$_{15}$ is thoroughly investigated in the present study, where we find that  a local distortion of the crystal structure plays an important role.

Speaking of superconductivity in the two-leg Hubbard ladder, the impact of incipient bands has been studied extensively in the past~\cite{KHA, Matsumoto,Matsumoto2,Kato,Sakamoto}. Namely, when the Fermi level intersects one of the bands (say, the bonding band), but barely touches the other (the antibonding band), namely, when the antibonding band is incipient, superconductivity is found to be strongly enhanced. Therefore, one may also expect such an effect to take place in the double-chain system, where ``incipient'' in this case means that, in the unfolded Brillouin zone, the Fermi level is located near, say, the local minimum of the band. We analyze the band structure and the mechanism of superconductivity in Pr$_2$Ba$_4$Cu$_7$O$_{15-\delta}$ in the present study, using fluctuation exchange (FLEX) approximation, and we indeed find that superconductivity can be strongly enhanced owing to the incipient band effect.

The paper is organized as follows.
Theoretical methods used in this study are described in Sec.~\ref{sec:methods}.
Sections~\ref{sec:DFT} and \ref{sec:dp} present first-principles and tight-binding analysis on the origin of a sizable $|t/t'|$ in Pr$_2$Ba$_4$Cu$_7$O$_{15}$.
After that, we show our FLEX analysis on the possible superconducting mechanism originating from a large $|t/t'|$ in Sec.~\ref{sec:SC}.
Section~\ref{sec:sum} summarizes this study.

\section{Methods\label{sec:methods}}

For density functional theory (DFT) calculations, we used the projector augmented wave (PAW) method~\cite{PAW} and the Perdew-Burke-Ernzerhof parametrization of the generalized gradient approximation~\cite{PBE} as implemented in the Vienna {\it ab initio} simulation package~\cite{VASP1,VASP2,VASP3,VASP4}. Core-electron states in PAW potentials were [Kr]$4d^{10}4f^{2}$, [Kr]$4d^{10}$, [Ar]$3d^{10}$, [Ar], and [He] for Pr, Ba, Sr, Cu, and O, respectively. Open-core treatment was applied to represent the Pr$^{3+}$ state.
We used a plane-wave cutoff energy of 520 eV for Kohn-Sham orbitals without including the spin-orbit coupling for simplicity. A $10\times 10\times 10$ ${\bm k}$-mesh was used for all DFT calculations.

We performed structural optimization until the Hellmann-Feynman force becomes less than 0.01 eV \AA$^{-1}$ for each atom. For this purpose, we only optimized atomic coordinates while fixing lattice constants extracted from experiments, which are listed on Table~\ref{tab:lattice_param}. For Pr$_2$Ba$_4$Cu$_7$O$_{15}$, we ignored the partial occupation of oxygen atomic sites representing randomness of the alignment direction of the single-chain structure, and assumed that the single chain aligns along the $a$ direction as the double chain does.
We also ignored the oxygen vacancy observed in experiments since its theoretical treatment is difficult.

\begin{table}
\caption{Lattice parameters used in this study, which were extracted from experimental studies.}
\label{tab:lattice_param}
\centering
\begin{tabular}{cccccc}
\hline
\hline
 & $a$ (\AA) & $b$ (\AA) & $c$ (\AA) & space group & Ref. \\
 \hline
  SrCu$_2$O$_3$ & 3.9313 & 11.5629 & 3.4926 & {\it Cmmm} & [\onlinecite{Sr123_strct}]\\
 PrBa$_2$Cu$_4$O$_8$ & 3.8837 & 3.90269 & 27.293 & {\it Ammm} & [\onlinecite{Pr124_strct}]\\
 Pr$_2$Ba$_4$Cu$_7$O$_{15}$ & 3.89 & 3.91 & 50.74 & {\it Ammm} & [\onlinecite{Sasaki}]\\
\hline
\hline
\end{tabular}
\end{table}

After DFT calculations, we extracted Wannier orbitals using Wannier90 software~\cite{wan1,wan2,wan3}.
To construct a $d_{x^2-y^2}$ model, we extracted all Cu-$d_{x^2-y^2}$ orbitals in the cell and Cu-$d_{3z^2-r^2}$ orbitals on the CuO$_2$ plane adjacent to the double chain to better reproduce the first-principles band structure. To construct a $dp$ model, we extracted all Cu-$d$ and O-$p$ orbitals in the cell.

To discuss superconductivity, we used a single-orbital Hubbard model, 
\begin{equation}
H = \sum_{i,j, \sigma} t_{ij} c_{i,\sigma}^{\dag} c_{j,\sigma} + U \sum_{i} n_{i, \uparrow} n_{i, \downarrow},\label{eq:Hubbard}
\end{equation}
where $i$ ($j$), $\sigma$, $t$, and $U$ denote a site in the unit cell, spin, hopping, and onsite Coulomb interaction, respectively,
and analyzed it within FLEX approximation~\cite{FLEX1, FLEX2}. 
Using the self energy calculated within FLEX where ring and ladder diagrams are considered, the linearized Eliashberg equation reads
\begin{align}
&\lambda \Delta({\bm k}, i\omega_n) \notag\\
&= -\frac{T}{N}\sum_{{\bm k'}, n'} \Gamma({\bm k} - {\bm k'}, i(\omega_n - \omega_{n'}))\notag\\
&\times G({\bm k'}, i\omega_{n'})
\Delta({\bm k'}, i\omega_{n'}) G(-{\bm k'}, -i\omega_{n'}),
\end{align}
where $T$, $N$, $\Delta$, $\Gamma$, $G$, $\lambda$ are the absolute temperature, the number of cells, the gap function, the pairing interaction, the renormalized Green's function, and the eigenvalue of the linearized Eliashberg equation, respectively.
We solved this equation to get $\lambda$ at a fixed temperature, which we regard as a quantity representing how high the superconducting critical temperature $T_{\mathrm{c}}$ of the system is.

\section{Results and Discussions\label{sec:res}}

\subsection{DFT band structure and hopping parameters\label{sec:DFT}}

\begin{figure*}
\begin{center}
\includegraphics[width=16 cm]{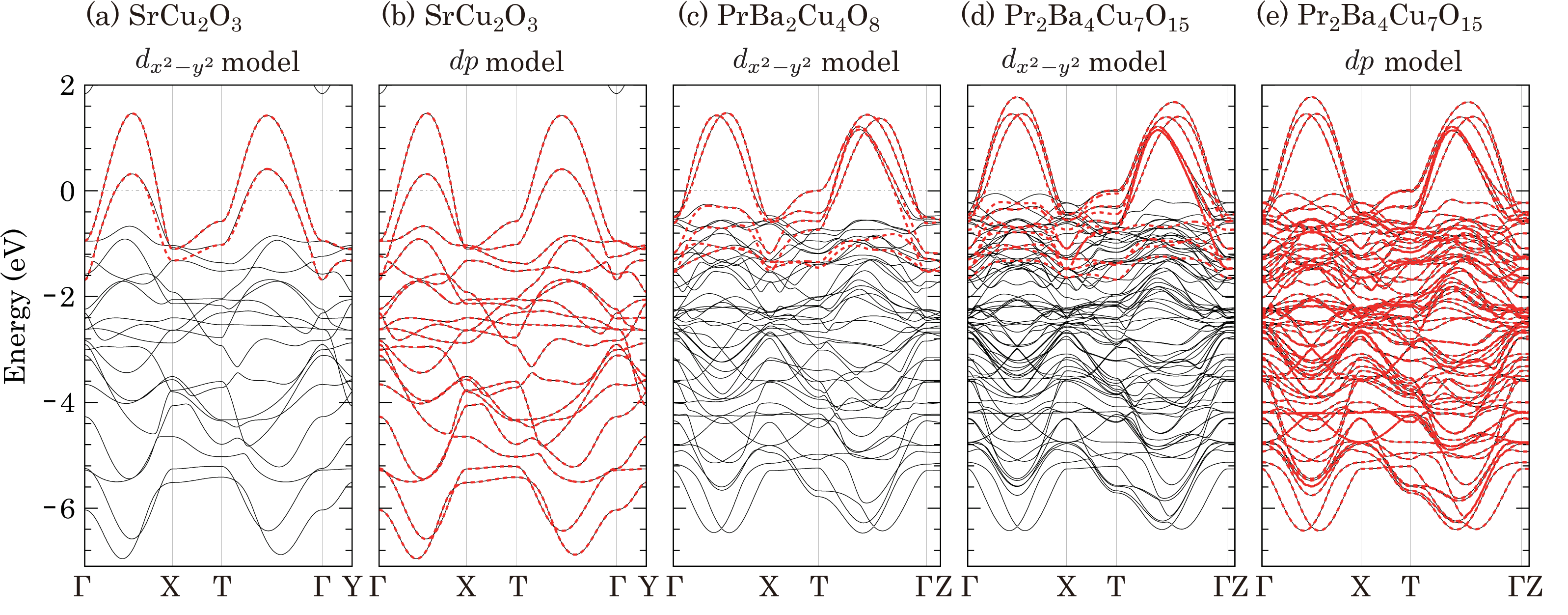}
\caption{Electronic band structures of (a)(b) SrCu$_2$O$_3$, (c) Pr$_2$Ba$_4$Cu$_7$O$_{15}$, and (d)(e) PrBa$_2$Cu$_4$O$_8$. First-principles band structure and that for the tight-binding model extracted by Wannierization are shown with black solid lines and red broken lines, respectively. The $d_{x^2-y^2}$ and $dp$ models are shown for (a)(c)(d) and (b)(e), respectively. The band dispersion was depicted along $\Gamma(0,0,0)$ - X$(2\pi/a,0,0)$ - T$(2\pi/a,0,2\pi/c)$ - $\Gamma$ - Y$(0,2\pi/b,0)$ for SrCu$_2$O$_3$ and $\Gamma(0,0,0)$ - X$(2\pi/a,0,0)$ - T$(2\pi/a,2\pi/b,0)$ - $\Gamma$ - Z$(0,0,2\pi/c)$ for Pr$_2$Ba$_4$Cu$_7$O$_{15}$ and PrBa$_2$Cu$_4$O$_8$ (in the cartesian coordinate).}
\label{fig:FPband}
\end{center}
\end{figure*}

Figure~\ref{fig:FPband} presents DFT band structures together with tight-binding band dispersions obtained by Wannierization.
Our tight-binding model well reproduces the original DFT band structure near the Fermi energy.
Extracted hopping parameters for the $d_{x^2-y^2}$ model are shown in Table~\ref{tab:hopping_param}.

\begin{figure}
\begin{center}
\includegraphics[width=6 cm]{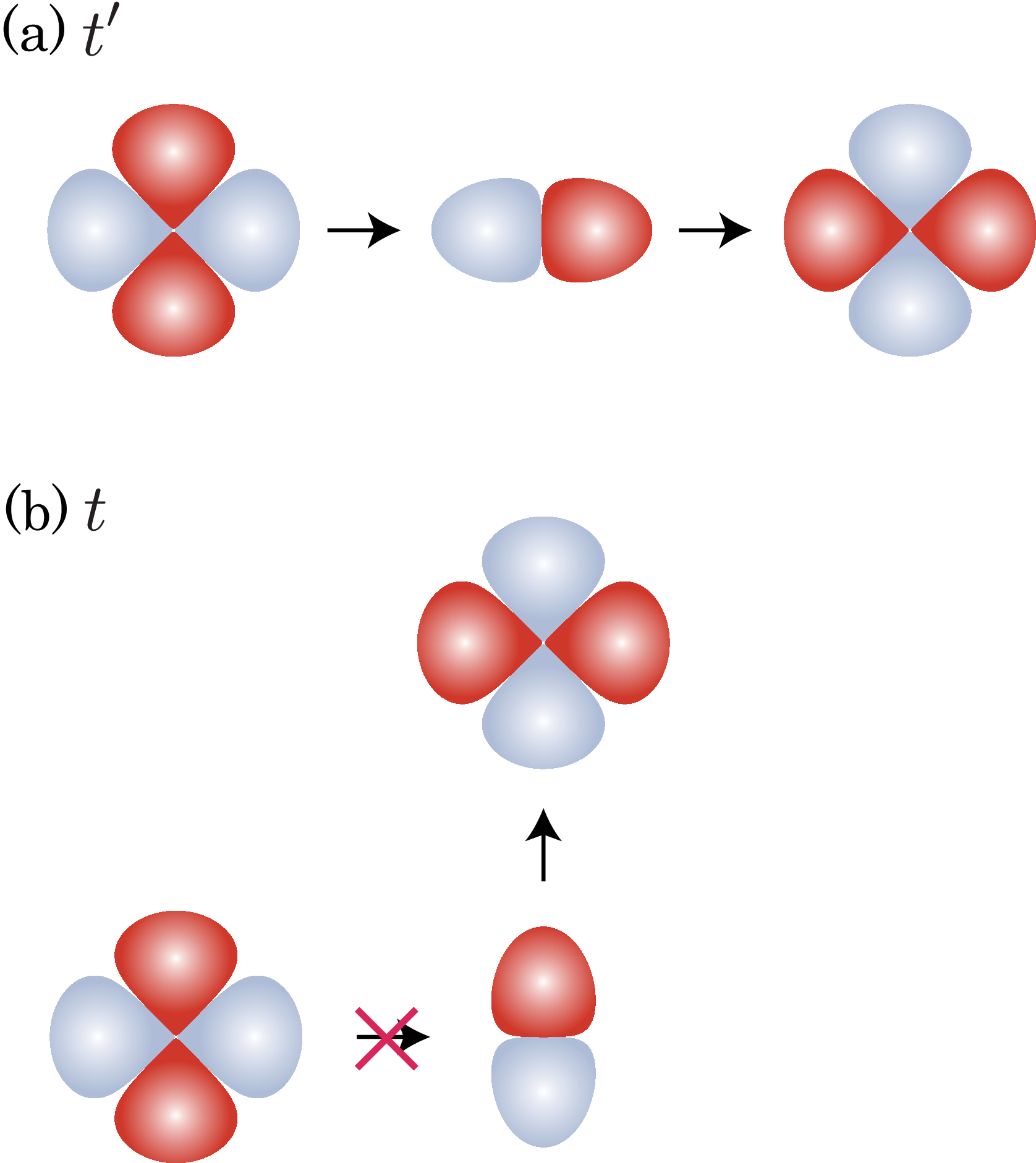}
\caption{Hopping processes (a) $t'$ and (b) $t$ between Cu-$d_{x^2-y^2}$ orbitals via an O-$p$ orbital in a double chain.}
\label{fig:cancel}
\end{center}
\end{figure}

It is noteworthy that $|t/t'|$ largely varies among these three materials:
$|t/t'|=0.06$, 0.23, and 0.22 for SrCu$_2$O$_3$, PrBa$_2$Cu$_4$O$_8$, and Pr$_2$Ba$_4$Cu$_7$O$_{15}$, respectively.
Considering the symmetry of atomic orbitals, it is rather natural to expect $t\sim 0$ as schematically depicted in Fig.~\ref{fig:cancel}.
As presented in Fig.~\ref{fig:cancel}(b), $d_{x^2-y^2}$-$p_y$ hopping along the $x$ axis is prohibited when $+y$ and $-y$ are equivalent for each atom, i.e., when the parity of $y\to -y$ can be locally defined.
While this symmetry does not rigorously hold in a double-chain structure, it is surprising to get a sizable $|t/t'|$ for PrBa$_2$Cu$_4$O$_8$ and Pr$_2$Ba$_4$Cu$_7$O$_{15}$.

In the following section, we investigate why such a large $|t/t'|$ is realized in these materials. Since the hopping parameters are similar between these two compounds, we shall focus on Pr$_2$Ba$_4$Cu$_7$O$_{15}$. We note that similar hopping parameters between PrBa$_2$Cu$_4$O$_8$ and Pr$_2$Ba$_4$Cu$_7$O$_{15}$ suggest that the absence of superconductivity in the former compound is not due to a change in the band structure.
In fact, some researchers pointed out that the absence of superconductivity in PrBa$_2$Cu$_4$O$_8$ might be due to a difference in the carrier density, which can be controlled in Pr$_2$Ba$_4$Cu$_7$O$_{15}$ through the oxygen deficiency in CuO single chains~\cite{Yamada}.

\begin{table}
\caption{Extracted hopping parameters (in eV) for the $d_{x^2-y^2}$ model.}
\label{tab:hopping_param}
\centering
\begin{tabular}{cccc}
\hline
\hline
 & $t$ & $t'$ & $|t/t'|$ \\
 \hline
 SrCu$_2$O$_3$ & \ 0.030 \ & \ $-0.480$ \  & \ 0.06 \ \\
 PrBa$_2$Cu$_4$O$_8$ & \ 0.120 \ & \ $-0.521$ \  & \ 0.23 \ \\
 Pr$_2$Ba$_4$Cu$_7$O$_{15}$ & \ 0.117 \  & \ $-0.522$ \  & \ 0.22 \ \\
\hline
\hline
\end{tabular}
\end{table}

\subsection{Origin of the sizable $|t/t'|$: Analysis on the $dp$ model\label{sec:dp}}

\subsubsection{Model simplification}

To investigate the origin of the sizable $|t/t'|$ in Pr$_2$Ba$_4$Cu$_7$O$_{15}$, we analyzed the $dp$ models for  SrCu$_2$O$_3$ and Pr$_2$Ba$_4$Cu$_7$O$_{15}$ and compared them.
Since the $dp$ model consists of all Cu-$d$ and O-$p$ orbitals in the cell, we first eliminated irrelevant orbitals from the model in the following way.

\begin{figure}
\begin{center}
\includegraphics[width=8.2 cm]{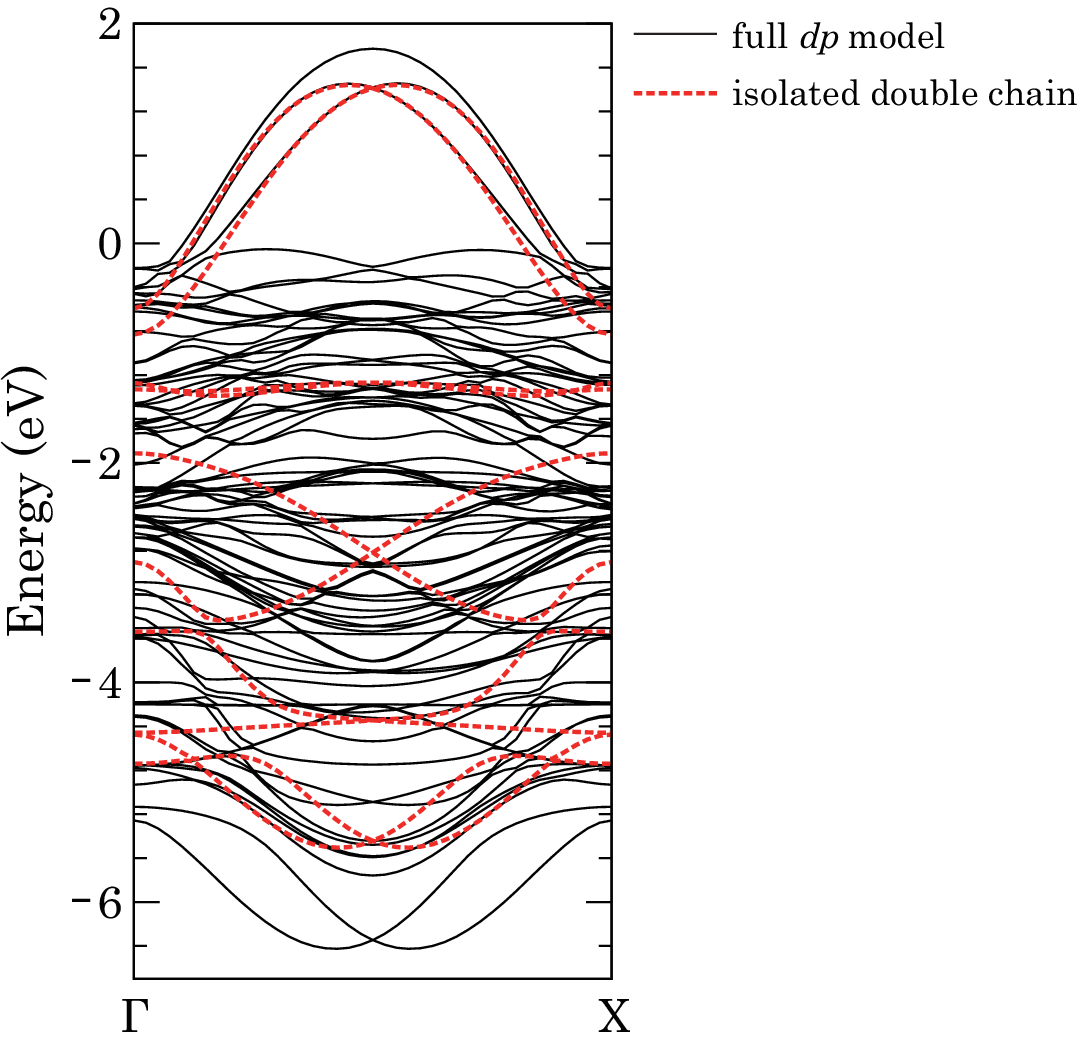}
\caption{Band dispersion for the tight-binding model of Pr$_2$Ba$_4$Cu$_7$O$_{15}$. Black solid and red broken lines present the band dispersion for the full $dp$ model and that for the `isolated double-chain' $dp$ model as defined in the main text, respectively. For the latter model, the band dispersion is slightly ($+0.14$ eV) shifted.}
\label{fig:band_full_vs_isolate_chain}
\end{center}
\end{figure}

Figure~\ref{fig:band_full_vs_isolate_chain} presents band dispersion for the full $dp$ model and that for the `isolated double-chain' $dp$ model of Pr$_2$Ba$_4$Cu$_7$O$_{15}$. 
Here, the latter model was constructed by extracting Cu-$d_{x^2-y^2}$ and O-$p_{x,y}$ orbitals inside the double chain [i.e., Cu-$d_{x^2-y^2}$ and O-$p_{x,y}$ orbitals defined on the atom sites shown in Fig.~\ref{fig:strct}(d)] from the full $dp$ model. Inter-double-chain hoppings were also eliminated, by which we call it the `isolated' double chain.
Since the band dispersion of the double chain is well reproduced by the isolated double-chain model, we focus on the latter model hereafter.

From here on, we consider a halved unit cell as shown in Fig.~\ref{fig:strct}(d), where O1-$p_{x,y}$, O2-$p_{x,y}$, and Cu-$d_{x^2-y^2}$ orbitals are included in the unit cell.
A neighboring unit cell is obtained by glide reflection of the cell (i.e., translation in addition to the $y\to -y$ reflection).
Thus, while this cannot be considered as a unit cell in first-principles calculations, this reduced cell can be regarded as a unit cell in model calculations. A similar model construction was applied to iron-based superconductors having nonsymmorphic symmetry~\cite{iron_theory}.

\begin{figure}
\begin{center}
\includegraphics[width=8.4 cm]{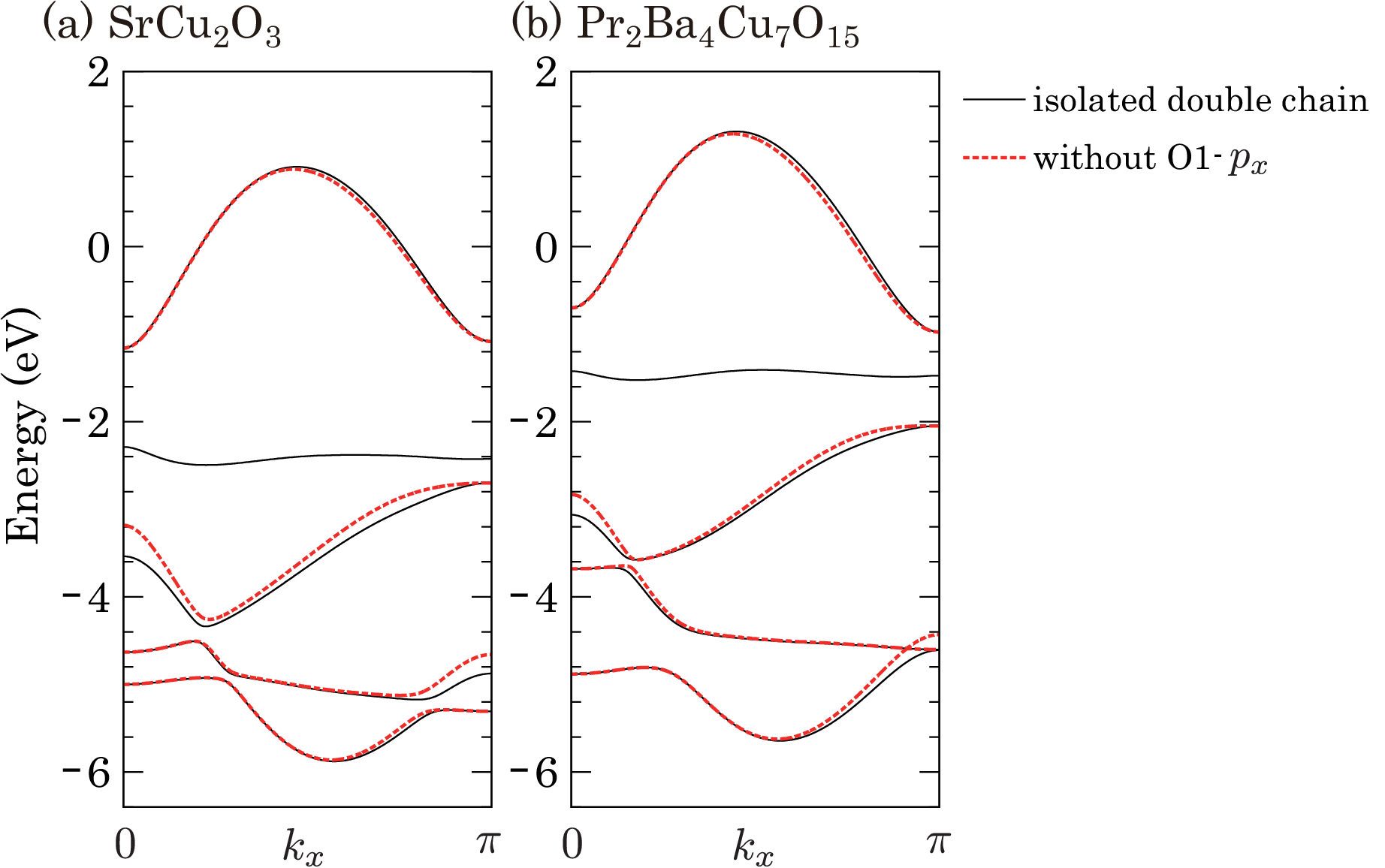}
\caption{Band dispersion for the tight-binding model of (a) SrCu$_2$O$_3$ and (b) Pr$_2$Ba$_4$Cu$_7$O$_{15}$. Black solid and red broken lines present the band dispersion for the `isolated double-chain' $dp$ model and that without the O1-$p_x$ orbital, respectively.}
\label{fig:band_compare_O1px}
\end{center}
\end{figure}

We applied further simplification to the model. Since we found that O1-$p_x$ has a negligible effect on the topmost $d_{x^2-y^2}$ band as shown in Fig.~\ref{fig:band_compare_O1px}, we also excluded O1-$p_x$ from the model in the following analysis. 
Note that the model has a single $d$ band in the folded Brillouin zone by halving the unit cell.

\subsubsection{Analysis of $|t/t'|$}

To theoretically pin down important hopping paths in the isolated double-chain $dp$ model, we evaluated how the effective value of $t$ changes when the model consists of a subset of orbitals.
We tried all possible combinations of the orbitals within the isolated double-chain model without O1-$p_x$, as listed in Table~\ref{tab:combination}, and extracted the hopping parameters $t$ and $t'$ by performing Fourier transformation for the topmost $d_{x^2-y^2}$ band for each model. Namely, considering the one-dimensional structure without inter-double-chain hoppings, the topmost $d_{x^2-y^2}$ band dispersion can be expanded as
\begin{equation}
\epsilon(k_x) = 2t \cos k_x + 2t' \cos (2k_x) + 2t'' \cos (3k_x) + \dots,
\end{equation}
where the coefficient of each term obtained by Fourier transform can be regarded as the effective hoppings for the $d_{x^2-y^2}$ model.

\begin{table}
\caption{All possible combinations of the orbitals for the isolated double-chain model without O1-$p_x$. Note that Cu-$d_{x^2-y^2}$ orbital is already taken because our purpose is to evaluate an effective hopping value $t$ between Cu-$d_{x^2-y^2}$.}
\label{tab:combination}
\centering
\begin{tabular}{ccccc}
\hline
\hline
 & Cu-$d_{x^2-y^2}$ & O1-$p_y$ & O2-$p_x$ & O2-$p_y$ \\
 \hline
model 1 & $\checkmark$ & & & \\
model 2 & $\checkmark$ & $\checkmark$ & & \\
model 3 & $\checkmark$ & & $\checkmark$ & \\
model 4 & $\checkmark$ & $\checkmark$ & $\checkmark$ & \\
model 5 & $\checkmark$ &  &  & $\checkmark$\\
model 6 & $\checkmark$ & $\checkmark$ & & $\checkmark$\\
model 7 & $\checkmark$ &  & $\checkmark$ & $\checkmark$\\
model 8 & $\checkmark$ & $\checkmark$ & $\checkmark$ & $\checkmark$\\
\hline
\hline
\end{tabular}
\end{table}

\begin{figure*}
\begin{center}
\includegraphics[width=16 cm]{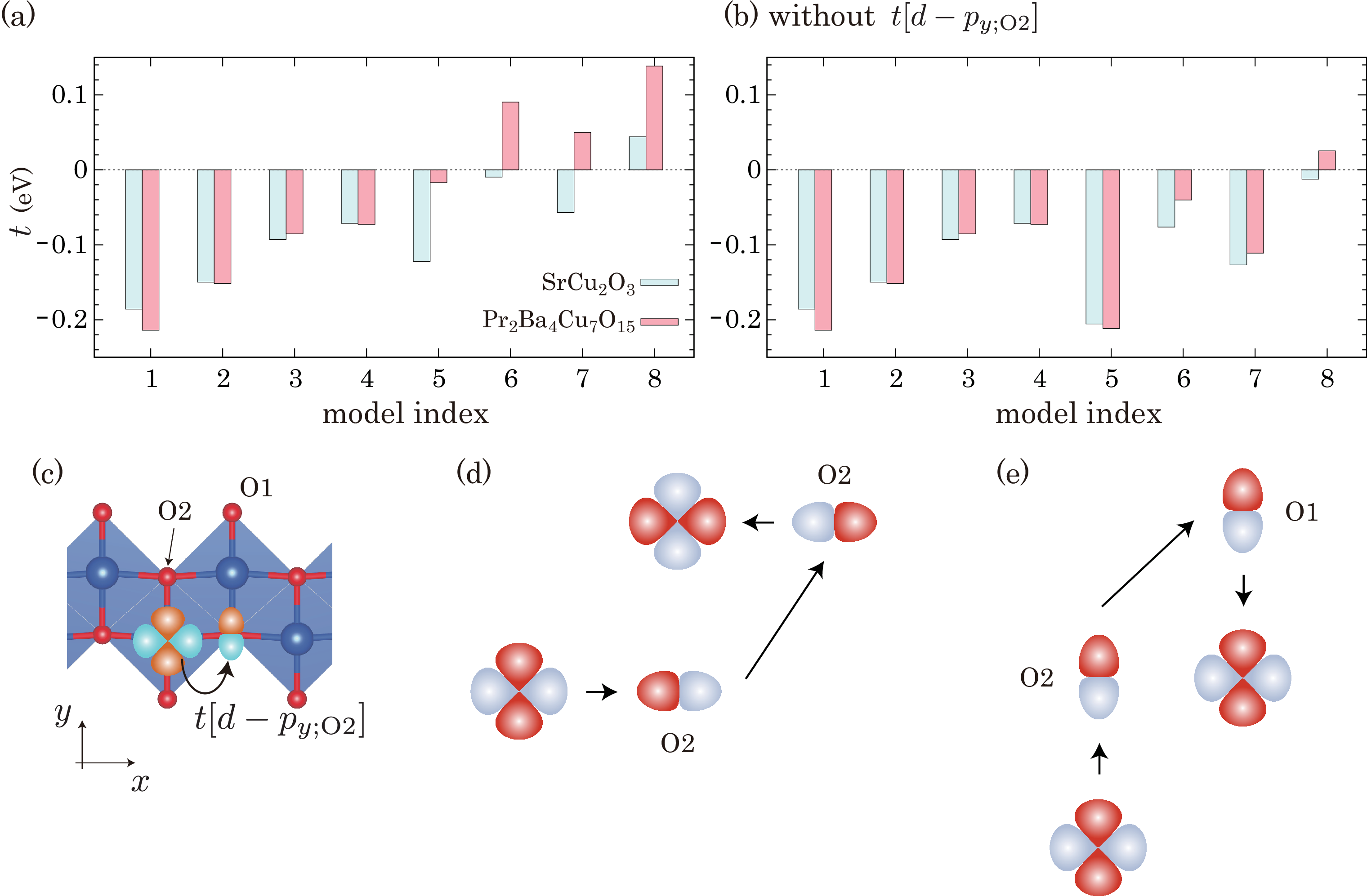}
\caption{(a) Extracted $t$ values for models defined in Table~\ref{tab:combination} for SrCu$_2$O$_3$ and Pr$_2$Ba$_4$Cu$_7$O$_{15}$. (b) Those evaluated by hypothetically excluding $t[d-p_{y;\mathrm{O2}}]$ from the models. (c) Definition of $t[d-p_{y;\mathrm{O2}}]$. (d)(e) Possible $d$-$p$-$p$-$d$ hopping paths contributing to $t$.}
\label{fig:hopping_models}
\end{center}
\end{figure*}

Figure~\ref{fig:hopping_models}(a) presents an extracted value of $t$ for each model, where the model index is defined in Table~\ref{tab:combination}.
We found that the extracted values of $t$ are roughly consistent between SrCu$_2$O$_3$ and Pr$_2$Ba$_4$Cu$_7$O$_{15}$ for models 1--4, but this is not the case for models 5--8.
Since model 5--8 contains O2-$p_y$ while it is not included in models 1--4, this result suggests that a main cause of the difference in $t$ between two materials is brought by O2-$p_y$.

To get further insight into the role of O2-$p_y$, in Fig.~\ref{fig:hopping_models}(b), we present $t$ evaluated for these models without including a specific hopping, $t[d-p_{y;\mathrm{O2}}]$, which is defined in Fig.~\ref{fig:hopping_models}(c). This is the nearest neighbor hopping between Cu-$d_{x^2-y^2}$ and O2-$p_y$ along the $x$ direction, which is prohibited when the $y\to -y$ symmetry exists for each atomic site and then $d_{x^2-y^2}$ and $p_y$ have even and odd parities, respectively. In fact, $t[d-p_{y;\mathrm{O2}}]$ in SrCu$_2$O$_3$ is small, 0.093 eV, even though the perfect $y\to -y$ symmetry no longer exists. On the other hand, we found that Pr$_2$Ba$_4$Cu$_7$O$_{15}$ has a sizable $t[d-p_{y;\mathrm{O2}}]=0.204$ eV. Since this hopping is expected to largely contribute to the $t$ hopping path shown in Fig.~\ref{fig:cancel}(b), the size of $t[d-p_{y;\mathrm{O2}}]$ can be the main origin of different $t$ between SrCu$_2$O$_3$ and Pr$_2$Ba$_4$Cu$_7$O$_{15}$; this expectation is verified in Fig.~\ref{fig:hopping_models}(b), where the difference between the two materials is drastically decreased compared with Fig.~\ref{fig:hopping_models}(a) by hypothetically excluding $t[d-p_{y;\mathrm{O2}}]$ from the models.

We note that, while the main difference in $t$ between SrCu$_2$O$_3$ and Pr$_2$Ba$_4$Cu$_7$O$_{15}$ is due to the size of $t[d-p_{y;\mathrm{O2}}]$, there are many other relevant hopping paths contributing to $t$.
In fact, the value of $t$ largely varies among different models in Fig.~\ref{fig:hopping_models}(b).
For example, a $d$-$p$-$p$-$d$ hopping path shown in Fig.~\ref{fig:hopping_models}(d) can contribute to $t$, which is included in models 3, 4, 7, and 8, where O2-$p_x$ is included.
Another $d$-$p$-$p$-$d$ path shown in Fig.~\ref{fig:hopping_models}(e) is included in models 6 and 8, where both O1-$p_y$ and O2-$p_y$ are included.
In Fig.~\ref{fig:hopping_models}(b), it seems that $t$ is increased by these hopping paths. 
As we have seen in the previous paragraph, $t$ is also increased by the hopping path using $t[d-p_{y;\mathrm{O2}}]$, which is most likely $d$-$p$-$d$.
On the other hand, the negative $t$ value for the model 1, where no oxygen orbitals are included, is the direct $d$-$d$ hopping.

To summarize, the direct $d$-$d$ hopping gives a negative $t \sim -0.2$ eV (see model 1 in Fig.~\ref{fig:hopping_models}(a) or (b)) while the $d$-$p$-$p$-$d$ (or longer) hopping paths via oxygen $p$-orbitals raise the value of $t$.
Although these two contributions are roughly canceled out (see model 8 in Fig.~\ref{fig:hopping_models}(b)), the $d$-$p$-$d$ hopping path with a large $t[d-p_{y;\mathrm{O2}}]$ results in a positive and sizable $t$ for Pr$_2$Ba$_4$Cu$_7$O$_{15}$ (see model 8 in Figs.~\ref{fig:hopping_models}(a)).

\begin{figure}
\begin{center}
\includegraphics[width=7 cm]{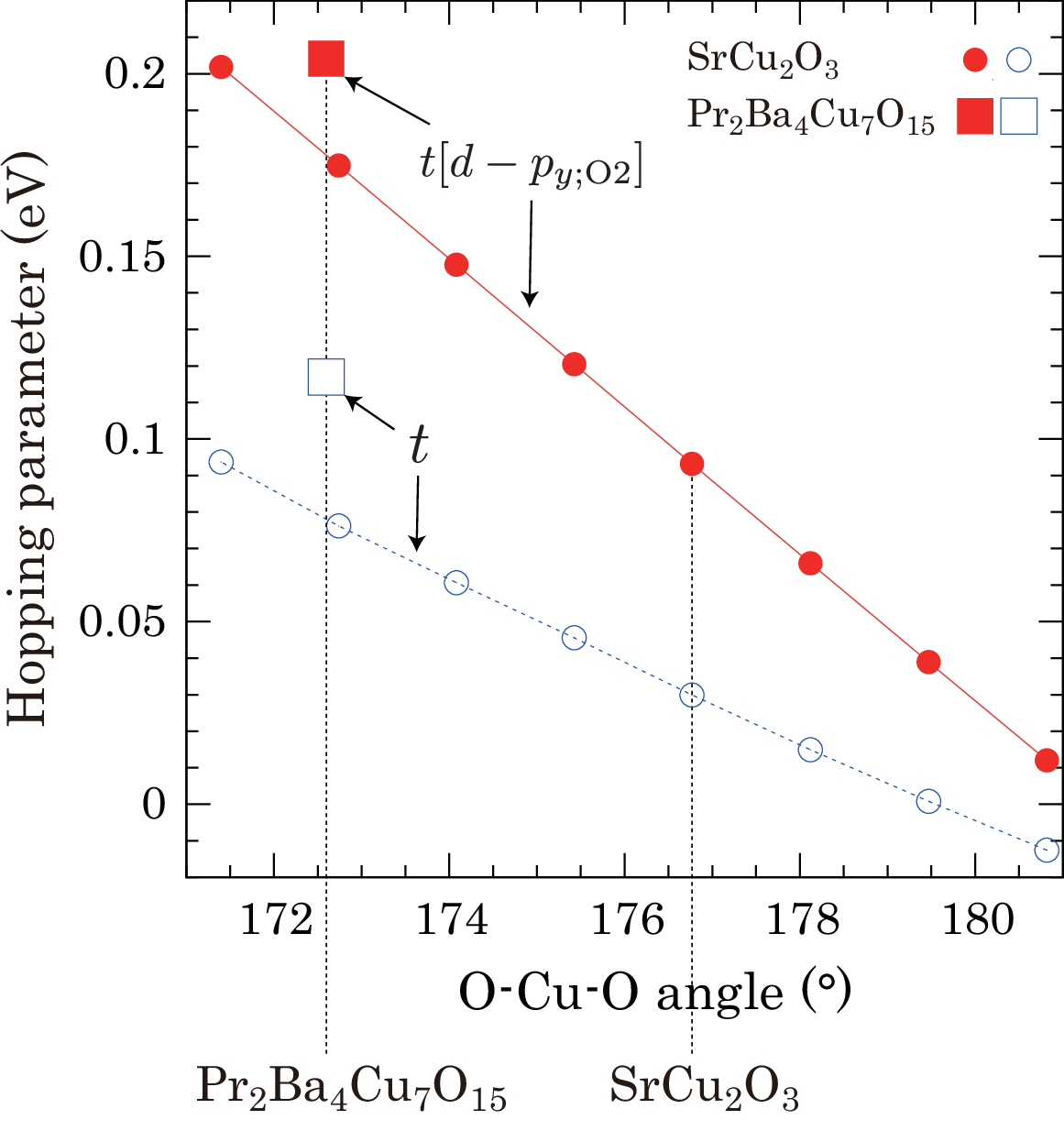}
\caption{Extracted $t$ (red filled symbols) and $t[d-p_{y;\mathrm{O2}}]$ (blue open symbols) values plotted against the O-Cu-O angle along the $x$ direction in the double chain. Square and circle symbols denote the values for Pr$_2$Ba$_4$Cu$_7$O$_{15}$ and SrCu$_2$O$_3$, respectively. For SrCu$_2$O$_3$, we hypothetically moved the O2 coordinate along the $y$ direction in calculation to change the O-Cu-O angle. Vertical black broken lines represent the Cu-O-Cu angles obtained by structural optimization for two compounds.}
\label{fig:hopping_angle}
\end{center}
\end{figure}

To investigate the reason for the sizable $t[d-p_{y;\mathrm{O2}}]$ in Pr$_2$Ba$_4$Cu$_7$O$_{15}$, we checked how the local structure of Cu-O double chains affects the hopping parameters.
For this purpose, we hypothetically changed the O2 coordinate along the $y$ direction for SrCu$_2$O$_3$ and extracted $t$ and $t[d-p_{y;\mathrm{O2}}]$ for each structure. The results are shown in Fig.~\ref{fig:hopping_angle}. Note that these $t$ values were simply extracted by Wannierization for the $d_{x^2-y^2}$ model.
This plot clearly shows that both $t$ and $t[d-p_{y;\mathrm{O2}}]$ are sizably increased when the O-Cu-O angle moves away from 180$^\circ$.
Therefore, a somewhat large discrepancy from 180$^\circ$ in the O-Cu-O angle for Pr$_2$Ba$_4$Cu$_7$O$_{15}$ seems to be important to understand the large $t[d-p_{y;\mathrm{O2}}]$ and therefore $t$ in this compound.

\subsection{Superconductivity in the double chain\label{sec:SC}}

\begin{figure*}
\begin{center}
\includegraphics[width=16 cm]{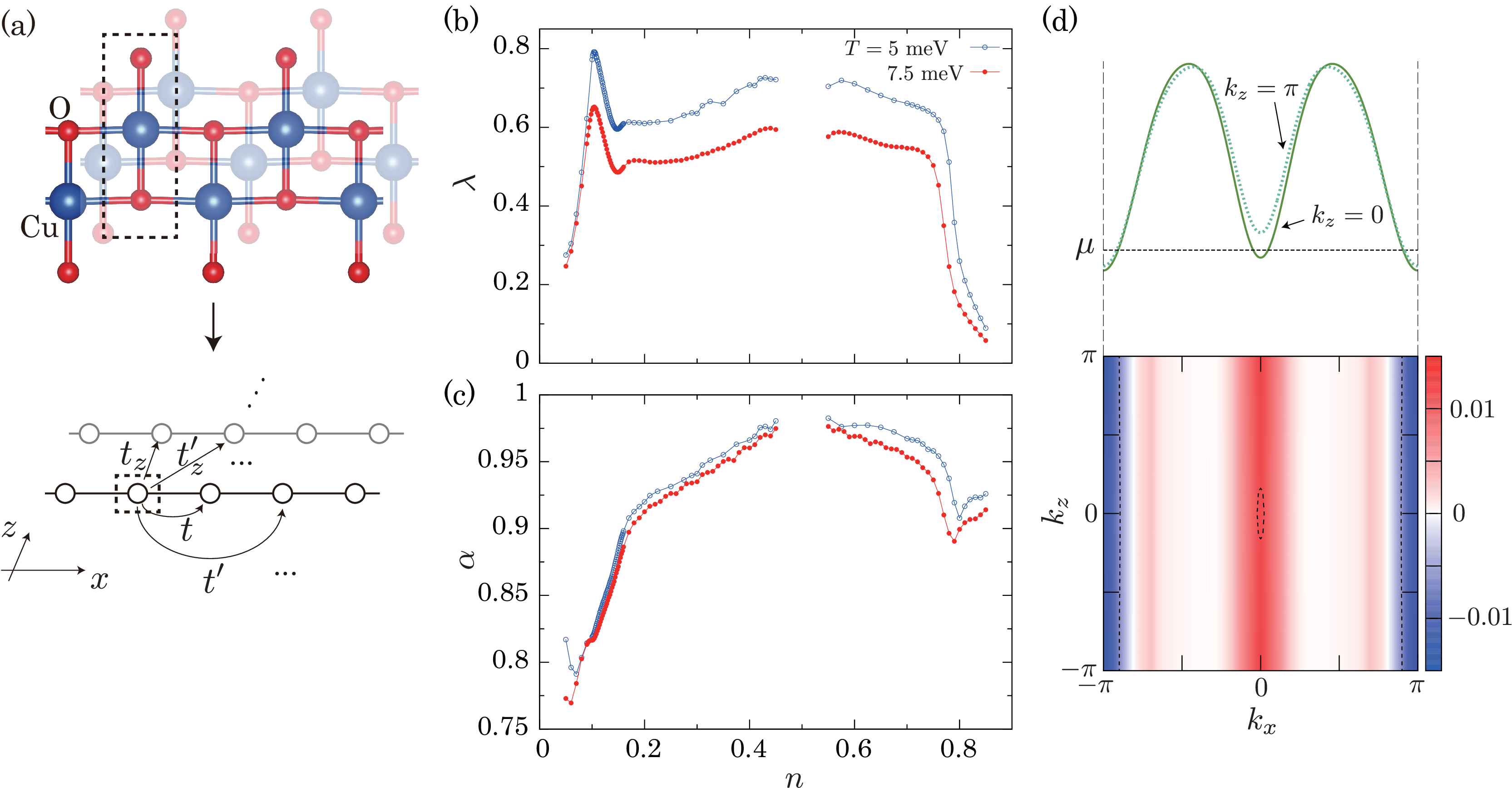}
\caption{Calculation results for FLEX $+$ Eliashberg calculations of the Cu-$d_{x^2-y^2}$ model for Pr$_2$Ba$_4$Cu$_7$O$_{15}$.
(a) A unit cell and the definition of hopping parameters, (b) the calculated eigenvalue of the linearized Eliashberg equation $\lambda$, (c) the Stoner factor $\alpha$, and (d) the gap function $\Delta({\bm k}, i\omega_0)$ at the lowest Matsubara frequency $\omega_0 = i\pi k_{\mathrm{B}}T$ (bottom) shown together with the tight-binding band dispersion (top) for $T=5$ meV, $n=0.103$, where $\lambda$ has a peak in (b). Broken black lines in the bottom panel presents the Fermi surface of the renormalized band dispersion (see the main text).}
\label{fig:real_mater_all}
\end{center}
\end{figure*}

We have investigated the origin of large $|t/t'|$ in Pr$_2$Ba$_4$Cu$_7$O$_{15}$.
In this section, we see how this aspect is related to the superconductivity in this compound.
For this purpose, we constructed a single-orbital Hubbard model consisting of the Cu-$d_{x^2-y^2}$ orbital on the double chain.
We extracted hopping parameters from the $d_{x^2-y^2}$ model as described before, and the parameters considered in the following analysis are shown in Table~\ref{tab:model_param}.
Here, the unit cell in model calculation is shown in Fig.~\ref{fig:real_mater_all}(a), which contains a single Cu site. 
Inter-double-chain hopping along the $z$ direction (the $a$ direction in the original lattice) was considered.

Then, we performed FLEX $+$ Eliashberg calculation using $U=2.5$ eV as a typical value of cuprates, the temperature $T=5$ and $7.5$ meV, a $256\times 16$ ${\bm k}$-mesh, and $2\times 2048$ Matsubara frequencies.
The obtained eigenvalue of the linearized Eliashberg equation $\lambda$ and the Stoner factor $\alpha=\mathrm{max}_{\bm q}[U\chi_0({\bm q},0)]$, where $\chi_0$ is the irreducible susceptibility at the lowest (Bosonic) Matsubara frequency in FLEX, are shown in Figs.~\ref{fig:real_mater_all}(b) and (c), respectively.
We note that it is difficult to theoretically determine the band filling $n$ for the double chain.
One reason for it is that a charge transfer among the CuO$_2$ plane, the double chain, and the single chain should occur in the actual material where the CuO$_2$ plane is a Mott insulator~\cite{Matsukawa,Yamada,Sasaki}. Another reason is that the actual material has oxygen vacancy~\cite{Matsukawa,Yamada}.
Thus, in this study, we set $n$ as a variable and plotted $\lambda$ and $\alpha$ as a function of $n$.
We excluded data points near the half filling $n=0.5$ since the strong electron correlation effects near the half filling is difficult to describe for the weak-coupling theory such as FLEX.

\begin{table}
\caption{Hopping parameters (eV) used in Fig.~\ref{fig:real_mater_all}, which were extracted by first-principles calculation and Wannier construction of the $d_{x^2-y^2}$ model.}
\label{tab:model_param}
\centering
\begin{tabular}{ccccccccc}
\hline
\hline
 & $t$ & $t'$ & $t''$ & $t'''$ & $t''''$ & $t_z$ & $t_z'$ & $t_z''$ \\
 \hline
 & 0.117 & $-0.522$ & $-0.041$ & $-0.069$ & $-0.012$ & $-0.013$ & $-0.014$ & $-0.014$ \\
\hline
\hline
\end{tabular}
\end{table}

We can see a notable peak of $\lambda$ around $n\sim 0.1$ in Fig.~\ref{fig:real_mater_all}(b), which was not discussed in the previous FLEX study~\cite{Nakano}.
It is also noteworthy that the Stoner factor $\alpha$ is less than 0.85 at the peak of $\lambda$, which is rather small for such a large $\lambda$ in FLEX $+$ Eliashberg calculations.
Such a small $\alpha$ indicates that the system is far from magnetic ordering, and suggests that a main glue of superconductivity is not a zero-energy spin fluctuation on the Fermi surface.
Figure~\ref{fig:real_mater_all}(d) presents the gap function $\Delta({\bm k}, i\omega_0)$ at the lowest (Fermion) Matsubara frequency $\omega_0 = i\pi k_{\mathrm{B}}T$ (bottom) shown together with the tight-binding band dispersion (top) for $T=5$ meV, $n=0.103$, where $\lambda$ has a peak. 
Broken black lines in the bottom panel presents the Fermi surface of the renormalized band dispersion: $\epsilon^0({\bm k}) + \mathrm{Re}[\Sigma({\bm k},i\omega_0)]$ where the self-energy correction was added to the non-interacting band dispersion $\epsilon^0({\bm k})$.
The gap function shows that the Cooper-pair scattering between $k_x=0$ and $k_x=\pi$ is a glue of the superconductivity.
While the Fermi surface is formed around $k_x=\pi$ irrespective of $k_z$, only a small pocket exists around the $k_x=0$ line.
Considering the small Stoner factor, the finite-energy spin fluctuation seems to be crucial here.
This situation reminds us the incipient-band superconductivity in multi-band systems, where the finite-energy pair scattering between the band intersecting the Fermi level and that lying just near the Fermi level is a key for superconductivity~\cite{KHA, Matsumoto,Matsumoto2,Kato,Sakamoto}. 
In the incipient-band mechanism, strong renormalization effects are alleviated by making one of the band edges incipient.
This aspect is advantageous, in particular when the band edge has a strong DOS peak such as (quasi-)one-dimensional systems~\cite{Matsumoto, Sakamoto} and systems with flat bands~\cite{Matsumoto,Aida}, while we have found the mechanism effectively works for other systems such as a bilayer square lattice~\cite{KHA,Matsumoto2,Kato}.
While our model for Pr247 has a single band, the band dispersion has a double-well structure with different energies for two valleys due to the sizable $|t/t'|$, by which the incipient-band mechanism can be realized.

Here we comment on the gap symmetry. In our model, the superconducting gap function is nodeless in each Fermi surface but changes its sign between large and small Fermi surfaces, which can be called $s^{\pm}$-wave. This is in sharp contrast to the usual one-dimensional chain where $t$ is dominant and the band dispersion is single-well. For a single-well band dispersion, the spin-singlet superconducting gap function should have a node on the Fermi surface, e.g., in a chain-like model~\cite{Ba213_prb} proposed for Ba$_2$CuO$_{3+\delta}$~\cite{Uchida}.

\begin{figure}
\begin{center}
\includegraphics[width=8 cm]{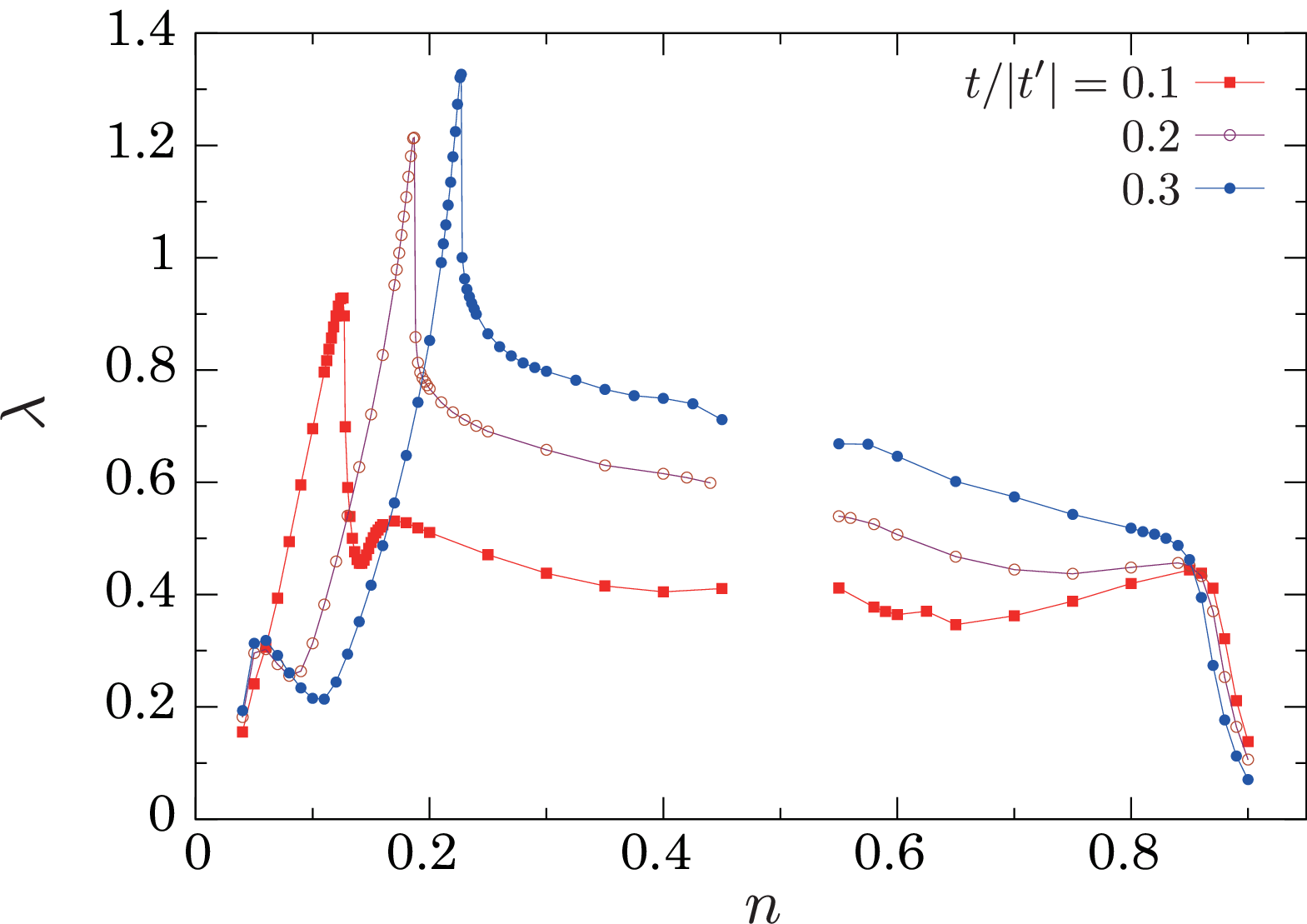}
\caption{FLEX $+$ Eliashberg calculation results for a simplified model. The calculated eigenvalue of the linearized Eliashberg equation $\lambda$ was plotted against the band filling $n$.}
\label{fig:lambda_model}
\end{center}
\end{figure}

A key in the superconducting mechanism described above is the double-well structure of the non-interacting band dispersion.
To understand this situation more clearly, we also performed FLEX $+$ Eliashberg calculation for a simplified model.
In the simplified model, we only considered $t$, $t'$, and $t_z$. The non-interacting band dispersion reads
\begin{equation}
\epsilon^0({\bm k}) = 2t \cos k_x + 2t' \cos (2 k_x) + 2t_z \cos k_z.
\end{equation}
The role of $t$ is clear: the $2t' \cos (2k_x)$ term makes the double-well band structure with the same energy for $k_x=0, \pi$ and the $2t \cos k_x$ term makes the energies of $k_x=0, \pi$ different.
We set $t'=-1$ as an energy unit, $t_z/|t'|=-0.05$, and tried several values of $t/|t'|$ to see the role of $t$.
While $t'$ was taken as an energy unit, we set its sign negative in accord with Table~\ref{tab:model_param}.
We used $U/|t'| = 5$, $T/|t'|=0.01$, a $256\times 16$ ${\bm k}$-mesh, and $2\times 4096$ Matsubara frequency.
The calculated results for $t/|t'|=0.1$, 0.2, 0.3 are shown in Fig.~\ref{fig:lambda_model}, where we can see similar peaks to that shown in Fig.~\ref{fig:real_mater_all}(b).
While the peak of $\lambda$ can be seen also for a relatively small $t/|t'|$, it is clear that a large $|t/t'|$ is desirable for superconductivity.
The peak position of $\lambda$ was shifted toward a larger band filling $n$ for a larger $t/|t'|$, which is naturally understood considering the incipient-band mechanism: the band filling where the Fermi level reaches the bottom of the band dispersion at $k_x=0$ becomes large for a large $t/|t'|$ [see Fig.~\ref{fig:real_mater_all}(d)].
In fact, we verified that the Fermi level locates around the bottom of the band dispersion at $k_x=0$ for all $t/|t'|$ when $\lambda$ reaches its peak.

\section{Summary\label{sec:sum}}

In this study, we have theoretically pinned down the origin of a large $|t/t'|$ in the double-chain structure of Pr$_2$Ba$_4$Cu$_7$O$_{15-\delta}$ using first-principles calculation and tight-binding-model analysis.
We have found that, in the nearest neighbor hopping $t$, $d$-$d$ and $d$-$p$-$p$-$d$ contributions roughly cancel each other out and the $d$-$p$-$d$  enhanced by the local distortion of the double chain is a key to get the large $|t|$. 
The enhancement of the $d$-$p$-$d$ hopping path is caused by $t[d-p_{y;\mathrm{O2}}]$, which is defined in Fig.~\ref{fig:hopping_models}(d) and becomes large for a non-linear Cu-O bond along the $x$ direction.
This is in contrast to the small interchain coupling in SrCu$_2$O$_3$, where the local distortion and thus the $d$-$p$-$d$ contribution are small.
Double-well band dispersion arising from the relatively large $|t/t'|$ allows the enhancement of spin-fluctuation-mediated superconductivity by the incipient-band mechanism, where the one band bottom plays a role of the incipient valley. Our study provides the important knowledge to understand the unique superconductivity in Pr$_2$Ba$_4$Cu$_7$O$_{15-\delta}$.

\acknowledgements
We thank Dr. Tatsuya Kaneko for fruitful discussions.
This study was supported by JSPS KAKENHI Grants No. JP22K04907 and No. JP24K01333.
The computing resource was supported by the supercomputer system in the Institute for Solid State Physics, the University of Tokyo.


\begin{thebibliography}{999}

\bibitem{Akimitsu}  M. Uehara, T. Nagata, J. Akimitsu, H. Takahashi, N. M{\^o}ri, and K. Kinoshita, Superconductivity in the Ladder Material Sr$_{0.4}$Ca$_{13.6}$Cu$_{24}$O$_{41.84}$, J. Phys. Soc. Jpn. {\bf 65}, 2764 (1996).

\bibitem{Hiroi} Z. Hiroi, M. Takano, M. Azuma, and Y. Takeda, A new family of copper oxide superconductors Sr$_{n+1}$Cu$_n$O$_{2n+1+\delta}$ stabilized at high pressure, Nature {\bf 364}, 315 (1993).

\bibitem{Uchida} W. M. Li, J. F. Zhao, L. P. Cao, Z. Hu, Q. Z. Huang, X. C. Wang, Y. Liu, G. Q. Zhao, J. Zhang, Q. Q. Liu, R. Z. Yu, Y. W. Long, H. Wu, H. J. Lin, C. T. Chen, Z. Li, Z. Z. Gong, Z. Guguchia, J. S. Kim, G. R. Stewart, Y.J. Uemura, S. Uchida, and C. Q. Jin, Superconductivity in a unique type of copper oxide, Proc. Natl. Acad. Sci. U.S.A. {\bf 116}, 12156 (2019).

\bibitem{Matsukawa} M. Matsukawa, Y. Yamada, M. Chiba, H. Ogasawara, T. Shibata, A. Matsushita, and Y. Takano, Superconductivity in Pr$_2$Ba$_4$Cu$_7$O$_{15-\delta}$ with metallic double chains, Physica C {\bf 411}, 101 (2004).

\bibitem{Yamada} Y. Yamada and A. Matsushita, Superconductivity in Pr$_2$Ba$_4$Cu$_7$O$_{15-\delta}$, Physica C {\bf 426}, 213 (2005).

\bibitem{Honnami} K. Honnami, M. Matsukawa, T. Senzaki, T. Toyama, H. Taniguchi, K. Ui, T. Sasaki, K. Takahashi, M. Hagiwara, Enhanced superconducting properties of double-chain based superconductor Pr$_2$Ba$_4$Cu$_7$O$_{15-\delta}$ synthesized by citrate pyrolysis technique, Physica C {\bf 585}, 1353869 (2021) .

\bibitem{Watanabe} S. Watanabe, Y. Yamada, and S. Sasaki, Cu-spin dynamics at zigzag chain of superconducting and nonsuperconducting Pr247 compounds, Physica C {\bf 426}, 473 (2005).

\bibitem{Toshima} S. Toshima, M. Matsukawa, T. Chiba, S. Kobayashi, S. Nimori, and M. Hagiwara, Magnetization and transport properties in the superconducting Pr$_2$Ba$_4$Cu$_7$O$_{15-\delta}$ with metallic double-chain, Physica C {\bf 480}, 1 (2012).

\bibitem{Sasaki} S. Nishioka, S. Sasaki, S. Nakagawa, M. Yashima, H. Mukuda, M. Yogi, and J. Shimoyama, Nuclear-spin evidence of insulating and antiferromagnetic state of CuO$_2$ planes in superconducting Pr$_2$Ba$_4$Cu$_7$O$_{15-\delta}$, Appl. Phys. Express {\bf 15}, 023001 (2022).

\bibitem{Sano} K. Sano, Y. \={O}no, and Y. Yamada, Superconductivity in the CuO Double Chain of Pr$_2$Ba$_4$Cu$_7$O$_{15-\delta}$ on the Basis of Tomonaga--Luttinger Liquid Theory, J. Phys. Soc. Jpn. {\bf 74}, 2885 (2005).

\bibitem{Nakano} T. Nakano, K. Kuroki, and S. Onari, Superconductivity due to spin fluctuations originating from multiple Fermi surfaces in the double chain superconductor Pr$_2$Ba$_4$Cu$_7$O$_{15-\delta}$, Phys. Rev. B {\bf 76}, 014515  (2007).

\bibitem{Sano2} K. Sano and Y. \={O}no, Superconductivity and Spin Gap in the Zigzag-Chain $t$-$J$ Model Simulating a CuO Double Chain in Pr$_2$Ba$_4$Cu$_7$O$_{15-\delta}$, J. Phys. Soc. Jpn. {\bf 76}, 113701 (2007).

\bibitem{Okunishi} K. Okunishi, Filling dependence of the zigzag Hubbard ladder for the quasi-one-dimensional superconductor Pr$_2$Ba$_4$Cu$_7$O$_{15-\delta}$, Phys. Rev. B {\bf 75}, 174514 (2007).

\bibitem{Berg} E. Berg, T. H. Geballe, and S. A. Kivelson, Superconductivity in zigzag CuO chains, Phys. Rev. B {\bf 76}, 214505 (2007).

\bibitem{Nishimoto} S. Nishimoto, K. Sano, and Y. Ohta, Phase diagram of the one-dimensional Hubbard model with next-nearest-neighbor hopping, Phys. Rev. B {\bf 77}, 085119  (2008).

\bibitem{Habaguchi} T. Habaguchi, Y. \={O}no, H. Ying Du Gh, K. Sano, and Y. Yamada, Electronic States and Superconducting Transition Temperature Based on the Tomonaga--Luttinger Liquid in Pr$_2$Ba$_4$Cu$_7$O$_{15-\delta}$, J. Phys. Soc. Jpn. {\bf 80}, 024708  (2011).

\bibitem{Kaneko} T. Kaneko, S. Ejima, K. Sugimoto, and K. Kuroki, Ground-State Properties of the $t$-$J$ Model for the CuO Double-Chain Structure, J. Phys. Soc. Jpn. {\bf 93}, 084703 (2024).

\bibitem{Fabrizio} M. Fabrizio, Superconductivity from doping a spin-liquid insulator: A simple one-dimensional example, Phys. Rev. B {\bf 54}, 10054 (1996).

\bibitem{KKArita} K. Kuroki, R. Arita, and H. Aoki, Numerical Study of a Superconductor-Insulator Transition in a Half-Filled Hubbard Chain with Distant Transfers, J. Phys. Soc. Jpn. {\bf 66}, 3371 (1997).

\bibitem{KHA} K. Kuroki, T. Higashida, and R. Arita, High-$T_c$ superconductivity due to coexisting wide and narrow bands: A fluctuation exchange study of the Hubbard ladder as a test case, Phys. Rev. B {\bf 72}, 212509 (2005).

\bibitem{Matsumoto} K. Matsumoto, D. Ogura, K. Kuroki, Wide applicability of high-$T_c$ pairing originating from coexisting wide and incipient narrow bands in quasi-one-dimensional systems, Phys. Rev. B {\bf 97}, 014516(2018).

\bibitem{Matsumoto2} K. Matsumoto, D. Ogura, K. Kuroki, Strongly Enhanced Superconductivity Due to Finite Energy Spin Fluctuations Induced by an Incipient Band: A FLEX Study on the Bilayer Hubbard Model with Vertical and Diagonal Interlayer Hoppings, J. Phys. Soc. Jpn. {\bf 89}, 044709 (2020).

\bibitem{Kato} D. Kato and K. Kuroki, Many-variable variational Monte Carlo study of superconductivity in two-band Hubbard models with an incipient band, Phys. Rev. Res. {\bf 2}, 023156 (2020).

\bibitem{Sakamoto} H. Sakamoto and K. Kuroki, Possible enhancement of superconductivity in ladder-type cuprates by longitudinal compression, Phys. Rev. Res. {\bf 2}, 022055(R) (2020).


\bibitem{PAW} G. Kresse and D. Joubert, From ultrasoft pseudopotentials to the projector augmented-wave method, Phys. Rev. B {\bf 59}, 1758 (1999).

\bibitem{PBE} J. P. Perdew, K. Burke, and M. Ernzerhof, Generalized Gradient Approximation Made Simple, Phys. Rev. Lett. {\bf 77}, 3865 (1996).

\bibitem{VASP1} G. Kresse and J. Hafner, $Ab$ $initio$ molecular dynamics for liquid metals, Phys. Rev. B {\bf 47}, 558 (1993).

\bibitem{VASP2} G. Kresse and J. Hafner, $Ab$ $initio$ molecular-dynamics simulation of the liquid-metal-amorphous-semiconductor transition in germanium, Phys. Rev. B {\bf 49}, 14251 (1994).

\bibitem{VASP3} G. Kresse and J. Furthm{\"u}ller, Efficiency of ab-initio total energy calculations for metals and semiconductors using a plane-wave basis set, Comput. Mater. Sci. {\bf 6}, 15 (1996).

\bibitem{VASP4} G. Kresse and J. Furthm{\"u}ller, Efficient iterative schemes for $ab$ $initio$ total-energy calculations using a plane-wave basis set, Phys. Rev. B {\bf 54}, 11169 (1996).

\bibitem{Pr124_strct} Y. Yamada, J. Ye, S. Horii, A. Matsushita, and S. Kubo, Crystal structure in PrBa$_2$Cu$_4$O$_8$ single crystals, J. Phys. Chem. Solids {\bf 62}, 191 (2001).

\bibitem{Sr123_strct} M. Azuma, H. Yoshida, T. Saito, T. Yamada, and M. Takano, Pressure-Induced Buckling of Spin Ladder in SrCu$_2$O$_3$, J. Am. Chem. Soc. {\bf 126}, 8244 (2004).

\bibitem{wan1} N. Marzari and D. Vanderbilt, Maximally localized generalized Wannier functions for composite energy bands, Phys. Rev. B {\bf 56}, 12847 (1997).

\bibitem{wan2} I. Souza, N. Marzari, and D. Vanderbilt, Maximally localized Wannier functions for entangled energy bands, Phys. Rev. B {\bf 65}, 035109 (2001).

\bibitem{wan3} G. Pizzi {\it et al}., Wannier90 as a community code: new features and applications, J. Phys.: Condens. Matter {\bf 32}, 165902 (2020).

\bibitem{FLEX1} N. E. Bickers, D. J. Scalapino, and S. R. White, Conserving Approximations for Strongly Correlated Electron Systems: Bethe-Salpeter Equation and Dynamics for the Two-Dimensional Hubbard Model, Phys. Rev. Lett. {\bf 62}, 961 (1989).

\bibitem{FLEX2} T. Dahm and L. Tewordt, Quasiparticle and Spin Excitation Spectra in the Normal and $d$-Wave Superconducting State of the Two-Dimensional Hubbard Model, Phys. Rev. Lett. {\bf 74}, 793 (1995).

\bibitem{VESTA} K. Momma and F. Izumi, {\it VESTA 3} for three-dimensional visualization of crystal, volumetric and morphology data, J. Appl. Crystallogr. \textbf{44}, 1272 (2011).

\bibitem{iron_theory} K. Kuroki, S. Onari, R. Arita, H. Usui, Y. Tanaka, H. Kontani, and H. Aoki, Unconventional Pairing Originating from the Disconnected Fermi Surfaces of Superconducting LaFeAsO$_{1-x}$F$_x$, Phys. Rev. Lett. {\bf 101}, 087004 (2008).

\bibitem{Aida} T. Aida, K. Matsumoto, D. Ogura, M. Ochi, and K. Kuroki, Theoretical study of spin-fluctuation-mediated superconductivity in two-dimensional Hubbard models with an incipient flat band, Phys. Rev. B {\bf 110}, 054516 (2024).

\bibitem{Ba213_prb} P. Adhikary, M. Gupta, A. Chauhan, S. Satpathy, S. Mukherjee, and B. R. K. Nanda, Unique ${d}_{xy}$ superconducting state in the cuprate ${\mathrm{Ba}}_{2}{\mathrm{CuO}}_{3.25}$, Phys. Rev. B {\bf 109}, L020505 (2024).


\end{thebibliography}
\end{document}